\documentclass[twocolumn]{aastex61}

\usepackage{amssymb}
\usepackage{bbm}
\usepackage{mathrsfs}
\usepackage{amsfonts}
\usepackage{mathptmx}
\usepackage{epsfig}
\usepackage{graphicx}
\usepackage{slashed}
\usepackage{color}
\usepackage{amsmath}

\begin{document}

\title{Determination of  plasma screening effects for thermonuclear reactions in laser-generated plasmas}

\author{Yuanbin \surname{Wu}}
\affil{Max-Planck-Institut f\"ur Kernphysik, Saupfercheckweg 1, D-69117 Heidelberg, Germany}
\email{yuanbin.wu@mpi-hd.mpg.de}

\author{Adriana \surname{P\'alffy}}
\affil{Max-Planck-Institut f\"ur Kernphysik, Saupfercheckweg 1, D-69117 Heidelberg, Germany}
\email{Palffy@mpi-hd.mpg.de}

\begin{abstract}

Due to screening effects, nuclear reactions in astrophysical plasmas may behave differently than in the laboratory. The possibility to determine the magnitude of these screening effects in colliding laser-generated plasmas is investigated theoretically, having as a starting point a proposed experimental setup with two laser beams at the Extreme Light Infrastructure facility. A laser pulse interacting with a solid target produces a plasma through the Target Normal Sheath Acceleration scheme, and this rapidly streaming plasma (ion flow) impacts on a secondary plasma created by the interaction of a second laser pulse on a gas jet target. We model this scenario here and calculate the reaction events for the astrophysically relevant reaction $^{13}$C($^4$He, $n$)$^{16}$O. We find that it should be experimentally possible to determine the plasma screening enhancement factor for fusion reactions by detecting the difference in reaction events between  two scenarios of ion flow interacting with the plasma target and a simple gas target. This provides a way to evaluate nuclear reaction cross-sections in stellar environments and can significantly advance the field of nuclear astrophysics.

\end{abstract}




\section{Introduction \label{sec:intr}}

In astrophysical environments, matter is usually found in the state of plasma. Stars are made of dense plasmas, and the intergalactic, interstellar, as well as the interplanetary medium consist mainly of diffuse plasmas. Thus, plasma plays a very important role in the universe. The properties of nuclear matter, such as reaction mechanisms and lifetimes, may drastically differ in the plasma environment from the ones in conventional laboratory environments. In this context, the role of charge screening is a crucial aspect.

Charge screening enhances the nuclear cross-sections by reducing the Coulomb barrier that reacting ions must overcome. There are basically two kinds of charge screening effects for the fusion of two positively charged nuclei \citep{AdelbergerRMP2011}. One is the electron screening applied to the laboratory reactions, which generally involve target nuclei bound in neutral atoms and molecules \citep{AssenbaumZPA1987}. The atomic (or molecular) electron cloud surrounding the target nucleus acts as a screening potential \citep{AssenbaumZPA1987}. This screening effect becomes important when the projectile energy is very low \citep{AdelbergerRMP2011, AssenbaumZPA1987}. The second type of screening effect is the plasma (electrostatic) screening \citep{SalpeterAJP1954, KellerAPJ1953, SalpeterAPJ1969, GraboskeAPJ1973, DzitkoAPJ1995, GruzinovAPJ1998} applied to the stellar (thermonuclear) reactions that happen in plasma environments in which most atoms are ionized. In plasmas, long-range electric fields are screened down by the dynamic flow of particles moving in response to electric fields. It is necessary to take into account the effects of this screening on the rate of thermonuclear reactions in stars \citep{AdelbergerRMP2011}.

In the past decades, many experiments that test the first type of electron screening in laboratory reactions with targets in the form of neutral atoms or molecules have been performed \citep{AdelbergerRMP2011, AssenbaumZPA1987, RolfsNIMPRSB1995, RolfsPPNP2001, EngstlerPLB1988, EngstlerZPA1992, GreifeZPA1995, AliottaNPA2001}. These experiments are not only important for the understanding of electron screening, but also for extracting bare cross-sections, which are needed as input for stellar reactions \citep{AdelbergerRMP2011}. In contrast to the case of laboratory screening, thus far, the plasma screening for stellar reactions is lacking in experimental tests \citep{AdelbergerRMP2011}. Determining the appropriate experimental conditions that allow us to evaluate the nuclear reactions in stellar environments could significantly contribute to the field of nuclear astrophysics. The study of direct measurements of reactions in laser-generated plasmas provides this opportunity \citep{RothELINP2015}. Appropriate experimental conditions for the study of atomic processes are also important for the understanding of the astrophysical observation results \citep{BernittN2012, OreshkinaPRL2014}; however it is out of the scope of the present work. We note that the use of colliding laser-generated plasma plumes  for nuclear physics studies was proposed a few years ago \citep{MascaliREDS2010} and was already experimentally demonstrated for the neutronless $p+ ^{11}\mathrm{Be}$ fusion reaction \citep{LabauneNC2013}.

\begin{figure}[h]
\begin{center}
\includegraphics[width=0.8\columnwidth]{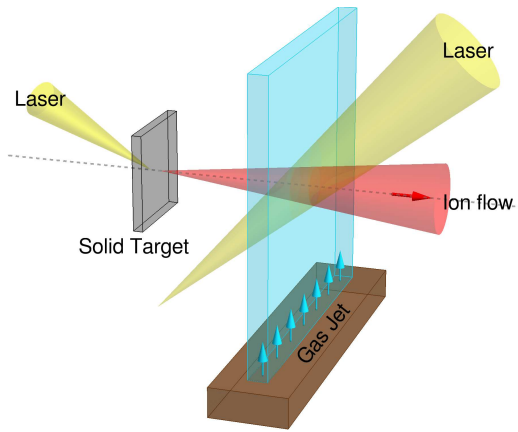}
\end{center}
\caption{Proposed experimental setup for nuclear reactions in laser-generated plasmas at ELI-NP for the $^{13}$C($^4$He, $n$)$^{16}$O reaction. The carbon ions are expelled and accelerated from the first solid target via the TNSA mechanism. They meet the gas jet He target (without the use of the second laser pulse) or the plasma He target and the nuclear reaction products are detected in the forward direction. A similar figure is illustrated in \cite{RothELINP2015}.} \label{fig:expsetup}
\end{figure}

At the upcoming Nuclear Pillar of the Extreme Light Infrastructure (ELI-NP) facility in construction in Romania, two high-power lasers working in parallel will soon be available \citep{RothELINP2015}. Each laser pulse can reach a peak power up to $10$ PW, with the typical pulse duration of $\sim 25$ fs. At this facility, an experimental setup in which two laser beams generate two colliding plasmas is envisaged in the Technical Design Report (TDR) \citep{RothELINP2015} [see also the published version of \cite{NegoitaRRP2016}]. This will give the opportunity to investigate nuclear reactions under plasma conditions. The target configuration of the proposed experimental setup is shown in Fig.~\ref{fig:expsetup}. A laser pulse interacting on a solid-state target produces a plasma through the Target Normal Sheath Acceleration (TNSA) mechanism [see \cite{MacchiRMP2013} and references therein], and this rapidly streaming plasma (ion flow) interacts with a secondary plasma created by the interaction of a second laser pulse on a gas jet target. Alternatively, without the use of the second laser pulse, this rapidly streaming plasma (ion flow) from TNSA interacts with a simple gas target. A high granularity SiC charged-particle detector and a neutron time-of-flight detector, not shown here in Fig.~\ref{fig:expsetup}, are placed after the target in the forward direction of the ion flow from TNSA, covering a quite large angle of the outgoing nuclear reaction products \citep{RothELINP2015}. This setup would allow the study of direct measurements of nuclear reaction rates in plasma environments. 

In this paper, we address this scenario from the theoretical side. We apply  an isothermal, fluid model \citep{MoraPRL2003, FuchsNP2006} to describe the TNSA mechanism in the laser-target interaction. With the ion spectrum obtained from this model, we study the nuclear fusion reactions in the interaction of the rapidly streaming plasma (ion flow) on the secondary target. As the first case of study, we analyze the reaction $^{13}$C($^4$He, $n$)$^{16}$O, which is one of the important helium burning processes as well as one of the main neutron sources for the $s$-process \citep{LooreBook1992, GallinoAPJ1998, HeilPRC2008}. We consider  the interaction of the TNSA-accelerated ions with both a plasma target and with a gas target. Our results show that a direct comparison of the neutron events considering the two types of targets should provide experimental access to the screening effects for the colliding plasma scenario. This would be the first time when experiments with laser-created plasmas in the laboratory could give insight on screening effects for thermonuclear reactions in astrophysical plasmas. We discuss the parameters that allow the determination of the plasma screening effect. 

The paper is organized as follows. In Sec.~\ref{sec:TNSA}, we present the fluid model for the TNSA scheme. We describe the plasma screening enhancement for fusion reactions of two positively charged nuclei in Sec.~\ref{sec:reactions}. Our results and discussion on the reaction $^{13}$C($^4$He, $n$)$^{16}$O follow in Sec.~\ref{sec:results}. We finally summarize and conclude in Sec.~\ref{sec:sumcon}. In Sec.~\ref{sec:TNSA}, SI units with $k_{\rm{B}} = 1$ are adopted, unless for some quantities the units are explicitly given. In Sec.~\ref{sec:reactions}, the Lorentz-Heaviside natural units $\hbar = c = \epsilon_0 = \mu_0 = k_{\rm{b}} = 1$ are adopted.

\section{TNSA model \label{sec:TNSA}}

In this section, we introduce a model to describe the TNSA scheme in the laser-target interaction, based on the fluid model of plasma expansion into a vacuum studied in \cite{MoraPRL2003}.  The model in \cite{MoraPRL2003} is a one-dimensional, isothermal, fluid model, in which the charge separation effects in the collisionless plasma expansion were studied. In this model, at time $t=0$, the plasma is assumed to occupy the half-space $x<0$. The ions are cold and initially at rest with density $n_{\rm{i}} = n_{\rm{i0}}$ for $x<0$ and $n_{\rm{i}} = 0$ for $x>0$ with a sharp boundary. The electron density $n_{\rm{e}}$ is continuous and corresponds to a Boltzmann distribution with the electron temperature $T_{\rm{e}}$. In addition, one also has $\Phi(-\infty) = 0$ and $n_{\rm{e0}} = Z_{\rm{i}} n_{\rm{i0}}$,  with $Z_{\rm{i}}$ being the ion charge state, $n_{\rm{e0}}$ the electron density in the unperturbed plasma, and $\Phi$ the electrostatic potential. Assuming quasi-neutrality in the expanding plasma, a self-similar expansion can be found for $x + c_{\rm{s}} t > 0$ \citep{MoraPRL2003},  i.e., $n_{\rm{e}} = Z_{\rm{i}} n_{\rm{i}} = n_{\rm{e0}} \exp{[-x/(c_{\rm{s}}t) - 1]}$ and $v_{\rm{i}} = c_{\rm{s}} + x/t$, where $c_{\rm{s}}$ is the ion sound speed, $c_{\rm{s}} = \sqrt{Z_{\rm{i}}T_{\rm{e}}/m_{\rm{i}}}$, $v_{\rm{i}}$ is the ion velocity, and $m_{\rm{i}}$ is the ion mass. Based on the self-similar expansion, and numerical solutions of the fluid model, Mora obtained the ion energy spectrum and the maximum ion energy \citep{MoraPRL2003}. 

Following the procedure described in \cite{FuchsNP2006}, we can apply this fluid model of plasma expansion to study the TNSA scheme in the laser-target interaction. Since this occurs in most experiments, \cite{FuchsNP2006} considers the possibility that a preformed plasma (preplasma) is present in front of the target owing to long-duration, low-level laser energy (prepulse) reaching the target before the main pulse. The electrons are accelerated following the irradiation by the main pulse. The density $n_{\rm{e0}}$ can be estimated by considering that the electrons accelerated at the target front surface are ballistically sprayed into the target. As the electrons are accelerated over the laser pulse duration and spread over the surface of the sheath $S_{\rm{sheath}}$, we have $n_{\rm{e0}} = N_{\rm{e}}/ (c \tau_{\rm{laser}} S_{\rm{sheath}})$, where $N_{\rm{e}}$ is the total number of electrons accelerated into the target, and $\tau_{\rm{laser}}$ is the laser pulse duration. The surface of the sheath $S_{\rm{sheath}}$ is given by 
\begin{equation}
   S_{\rm{sheath}} = \pi (r_0 + d_{\rm{t}} \tan{\theta})^2,
\end{equation}
depending on the half-angle divergence ($\theta \sim 25^{\rm{o}}$) of the hot electron inside the target \citep{FuchsPRL2003, FuchsNP2006}, the target thickness $d_{\rm{t}}$ and the initial radius $r_0$ of the zone over which the electrons are accelerated at the target front surface, i.e., the laser spot [$r_0$ is given by half of the full width at half maximum (FWHM)].

According to the model in \cite{MoraPRL2003}, the maximum (cutoff) energy that can be gained by the accelerated ions is \citep{FuchsNP2006}
\begin{equation} \label{eq:enmax}
   \mathbb{E}_{\rm{max}} = 2 \mathcal{E}_0 \left[ \ln\left(t_{\rm{p}} + \sqrt{t_{\rm{p}}^2 + 1}\right) \right]^2,
\end{equation}
where $t_{\rm{p}} = \omega_{\rm{pi}} t_{\rm{acc}}/\sqrt{2 \rm{Exp}}$ is the normalized acceleration time with ${\rm{Exp}}$ denoting the exponential constant that is the base of the natural logarithm, and $\mathcal{E}_0 = Z_{\rm{i}} T_{\rm{e}}$. Furthermore, $\omega_{\rm{pi}} $ is the ion plasma frequency, $\omega_{\rm{pi}} = \sqrt{n_{\rm{e0}} Z_{\rm{i}} e^2/(m_{\rm{i}}\epsilon_0)}$, and $t_{\rm{acc}}$ is  the effective acceleration time. As shown in \cite{FuchsNP2006}, $t_{\rm{acc}} \sim 1.3 \tau_{\rm{laser}}$ matches well with experimental results. This scaling provides similar results [see \cite{FuchsNP2006}] to the one presented in \cite{MoraPRE2005}, where a more refined fluid model of plasma expansion into a vacuum was introduced, which takes into account the finite size of the plasma. The number of accelerated ions per unit energy is given by \citep{MoraPRL2003, FuchsNP2006}
\begin{equation} \label{eq:enspe}
   \frac{d N}{d\mathbb{E}} = \frac{n_{\rm{i0}} c_{\rm{s}} t_{\rm{acc}} S_{\rm{sheath}}}{\sqrt{2\mathbb{E} \mathcal{E}_0}} \exp{\left(-\sqrt{2\mathbb{E}/ \mathcal{E}_0}\right)}.
\end{equation}

In order to obtain the cutoff energy and the energy spectrum of the ions, the ion and electron densities and the electron temperature must be known. The ponderomotive scaling \citep{WilksPRL1992, MalkaPRL1996, FuchsNP2006} is adopted for the electron temperature, 
\begin{equation} \label{eq:tepon}
   T_{\rm{e}} = m_{\rm{e}} c^2 \left[ \sqrt{1+ I \lambda_{\rm{\mu m}}^2/(1.37\times 10^{18})}  - 1 \right] ,
\end{equation}
where $m_{\rm{e}}$ is the electron mass, $I$ is the laser intensity in $\rm{W/cm^{2}}$, and $\lambda_{\rm{\mu m}}$ is the laser wavelength in micrometers. As shown in \cite{FuchsNP2006}, the total number of electrons accelerated into the target is
\begin{equation} \label{eq:tne}
   N_{\rm{e}} = f_{\rm{abs}} E_{\rm{laser}}/T_{\rm{e}},
\end{equation} 
where $E_{\rm{laser}}$ is the laser energy and $f_{\rm{abs}}$ is the fraction of laser light that is absorbed into the preplasma as fast electrons. According to experimental results, the empirical fraction $f_{\rm{abs}}$ can be given as \citep{FuchsNP2006}
\begin{equation} \label{eq:empfabs}
   f_{\rm{abs}} = 1.2 \times 10^{-15} I^{0.74}
\end{equation}
with a maximum of 0.5. Alternatively, one can use more recent experimental laser absorption data that has been reported for a
large range of intensities \citep{PingPRL2008, LevyNC2014}.  We note that the agreement with the empirical expression [Eq.~(\ref{eq:empfabs})] used above is good for the laser parameter regime of TNSA under consideration, when taking into account the experimental error bars.

We now turn to the charge state of the ions in the plasma. As shown in \cite{HegelichPRL2002}, with the strong space-charge field created at the rear surface of the target in TNSA, field ionization by barrier suppression is the dominant ionization mechanism at the rear of the target. The $q+$ ionic state will be created as soon as the electric field is above the threshold \citep{HegelichPRL2002} 
\begin{equation}
  E_{q}^{\rm{thresh}} = U_{q}^2 \epsilon_0 \pi/(qe),
\end{equation}
where $U_{q}$ is the ionization potential in eV. As a first approximation, we assume $Z_{\rm{i}}$ to be the charge state that can be reached under the initial electric field at the ion front predicted by the fluid model \citep{MoraPRL2003} discussed here,
\begin{equation}
  E_{\rm{front,0}} = \sqrt{2/\rm{Exp}} (n_{\rm{e0}} T_{\rm{e}} /  \epsilon_0)^{1/2}.
\end{equation}
For the case of $^{13}$C under consideration, we adopt the values of $U_{q}$ and $E_{q}^{\rm{thresh}}$ used in \cite{HegelichPRL2002}. These values are also shown in Table \ref{tabethresh}.

\begin{table}[h]  \addtolength{\tabcolsep}{12pt}
\begin{center}
\begin{tabular}{ccc}
  \hline\hline
  $q$  & $U_{q}$ [eV] & $E_{q}^{\rm{thresh}}$ [$\rm{V/m}$]  \\
  \hline
  $1$  & $11.2$           & $2.2 \times 10^{10}$ \\
  $2$  & $24.4$           & $5.2 \times 10^{10}$ \\ 
  $3$  & $47.9$           & $1.3 \times 10^{11}$ \\ 
  $4$  & $64.5$           & $1.8 \times 10^{11}$ \\ 
  $5$  & $392$            & $5.3 \times 10^{12}$ \\ 
  $6$  & $490$            & $7.0 \times 10^{12}$ \\   
  \hline\hline
\end{tabular}
\end{center}
\caption{The ionization potential $U_q$ and threshold $E_{q}^{\rm{thresh}}$ of the electric field for the creation of the $q+$ charged state state of carbon. These values are adopted from \cite{HegelichPRL2002}.}
\label{tabethresh}
\end{table}

\section{Nuclear reactions in plasmas \label{sec:reactions}}

The cross-section for the fusion of two positively charged nuclei is often written as [see, e.g., \cite{AssenbaumZPA1987, AtzeniBook2004, RaiolaEPA2002}]
\begin{equation} \label{eq:csnr}
   \sigma(\mathbb{E}_{\rm{c}}) = \frac{S(\mathbb{E}_{\rm{c}})}{\mathbb{E}_{\rm{c}}} \exp{(-2\pi \eta)},
\end{equation}
where $\mathbb{E}_{\rm{c}}$ is the center-of-mass (c.m.) energy, $S(\mathbb{E}_{\rm{c}})$ is the astrophysical factor, and $\eta$ is the Sommerfeld parameter, $\eta = Z_1 Z_2 \alpha \sqrt{m_{\rm{r}}/(2\mathbb{E}_{\rm{c}})}$ with $\alpha$ being the fine-structure constant. Here, $Z_1$ and $Z_2$ are the charge numbers of the interacting nuclei and $m_{\rm{r}}$ is their reduced mass, $m_{\rm{r}} = m_1 m_2/(m_1 + m_2)$, where $m_1$ and $m_2$ are the masses of the nuclei. The cross-section provided in Eq.~(\ref{eq:csnr}) assumes that the Coulomb potential of the interacting nuclei is that resulting from bare nuclei. However, this picture of nuclear reactions could be drastically affected inside plasmas, owing to the charge screening effects. According to the classical theory of plasmas, in weakly coupled plasmas, i.e., plasmas in which the Coulomb interaction energy between the nucleus and the nearest few electrons and nuclei is small compared to the thermal energy $T$, each particle only feels the effect of the particles at a distance smaller than the Debye length while, on longer scales, the plasma is quasi-neutral \citep{AtzeniBook2004}. The nuclear reaction rate in such a plasma is therefore enhanced by a factor of \citep{SalpeterAJP1954, GruzinovAPJ1998}
\begin{equation} \label{eq:salscr}
   g_{\rm{scr}} = \exp{\left( \frac{Z_1 Z_2 \alpha}{T \lambda_{\rm{D}}} \right)},
\end{equation}
where $\lambda_{\rm{D}}$ is the Debye length,
\begin{equation} \label{eq:debl}
   \frac{1}{\lambda_{\rm{D}}^2} = 4\pi n \alpha \zeta^2/T, 
\end{equation}
with
\begin{equation} \label{eq:zt}
   \zeta = \left[ \Sigma_i X_i \frac{Z_i^2}{A_i} + \left( \frac{f'}{f} \right) \Sigma_i X_i \frac{Z_i}{A_i} \right]^{1/2},
\end{equation}
and $n$ being the baryon number density. Here, $X_i$, $Z_i$, and $A_i$ are the mass fraction, the nuclear charge, and the atomic weight of ions of type $i$, respectively. The ratio $f'/f$ accounts for the electron degeneracy \citep{SalpeterAJP1954}, which for the plasmas under consideration $f'/f \simeq 1$. The detailed description of the term $f'/f$ can be found in \cite{SalpeterAJP1954}. It is important to notice here that the Salpeter formula (\ref{eq:salscr}) is valid under the assumption \citep{SalpeterAJP1954}
\begin{equation}
   \mathcal{I}_{\rm{z}} = \left( \frac{Z\alpha}{2a_{\rm{0z}}} \right) T^{-1} \ll 1,
\end{equation}
where $\mathcal{I}_{\rm{z}}$ stands for the ratio of the ionization potential of a $K$-shell electron in a hydrogen-like atom of charge $Z$ to the mean thermal energy and  $a_{\rm{0z}}$ is the Bohr radius for such an atom. Under this assumption, all atoms are completely ionized.

The averaged reactivity for the fusion of two positively charged nuclei in plasmas can be written as \citep{SalpeterAJP1954, GruzinovAPJ1998, AtzeniBook2004}
\begin{equation}
   <\!\!\!\sigma v\!\!\!>_{\rm{scr}} = g_{\rm{scr}}  <\!\!\!\sigma v\!\!\!>,
\end{equation}
where $<\!\!\!\sigma v\!\!\!>$ is the averaged reactivity neglecting screening.

The expression (\ref{eq:salscr}) is valid for the weak-screening case. In astrophysical environments, in addition to the weak-screening regime, it is also possible to reach the intermediate-screening and the strong-screening regimes \citep{SalpeterAJP1954, SalpeterAPJ1969, GraboskeAPJ1973, DzitkoAPJ1995, GruzinovAPJ1998} when going to more extreme conditions. However, in the plasmas under consideration here, which are produced by an intense optical laser interacting on a gas jet target, the density is typically lower than $10^{21}$ cm$^{-3}$ and the temperature reaches values between $\sim 100$ eV and a few keV. For these parameters, the weak-screening effect applies and we focus on this regime in the following.

\section{Results and discussion \label{sec:results}} 

In this section, we present our numerical results and a discussion on the reaction $^{13}$C($^4$He, $n$)$^{16}$O in $^4$He plasmas. As shown in the Salpeter formula (\ref{eq:salscr}), high-density plasmas are preferred in order to have a significantly large screening enhancement. With already available technology, gas jets with densities of several times of $10^{20}$ cm$^{-3}$ or even $10^{21}$ cm$^{-3}$ can be obtained [see, e.g., \cite{SchmidRSI2012, SyllaRSI2012}]. Therefore, in the present work, we focus on the density range between $10^{20}$ cm$^{-3}$ and $10^{21}$ cm$^{-3}$ for $^4$He. The plasma target is then created by a second laser interacting with the gas jet, as illustrated in Fig.~\ref{fig:expsetup}. The requirements on this laser are less stringent than the ones for the first laser responsible for TNSA; in particular, a large power and high-energy pulse would be advantageous for a large focal spot and stable operation over a longer pulse duration. However, intensities of $10^{14}$ W/cm$^2$ are already sufficient to create the target plasma. We note that some rough numerical simulations for nuclear reactions in the gas jet target have been presented in the original ELI-NP TDR \citep{RothELINP2015}, however, focusing mostly on low gas jet densities ($\sim 10^{18}$ cm$^{-3}$) and the low-energy part of the reactions. In this work, we go beyond that to address the plasma screening effects in fusion reactions.

\begin{figure}[h]
\begin{center}
\includegraphics[width=0.92\columnwidth]{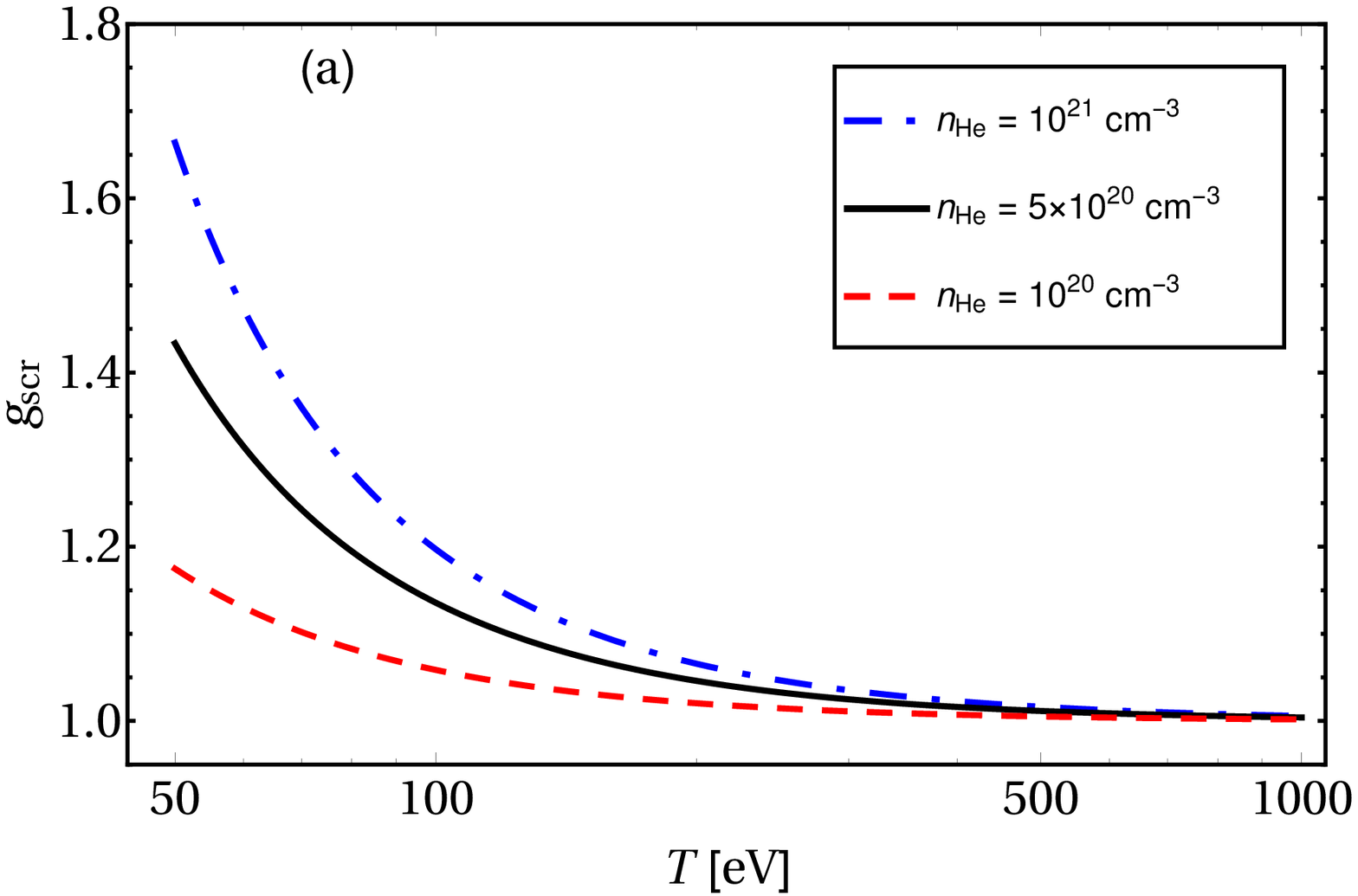}
\includegraphics[width=0.92\columnwidth]{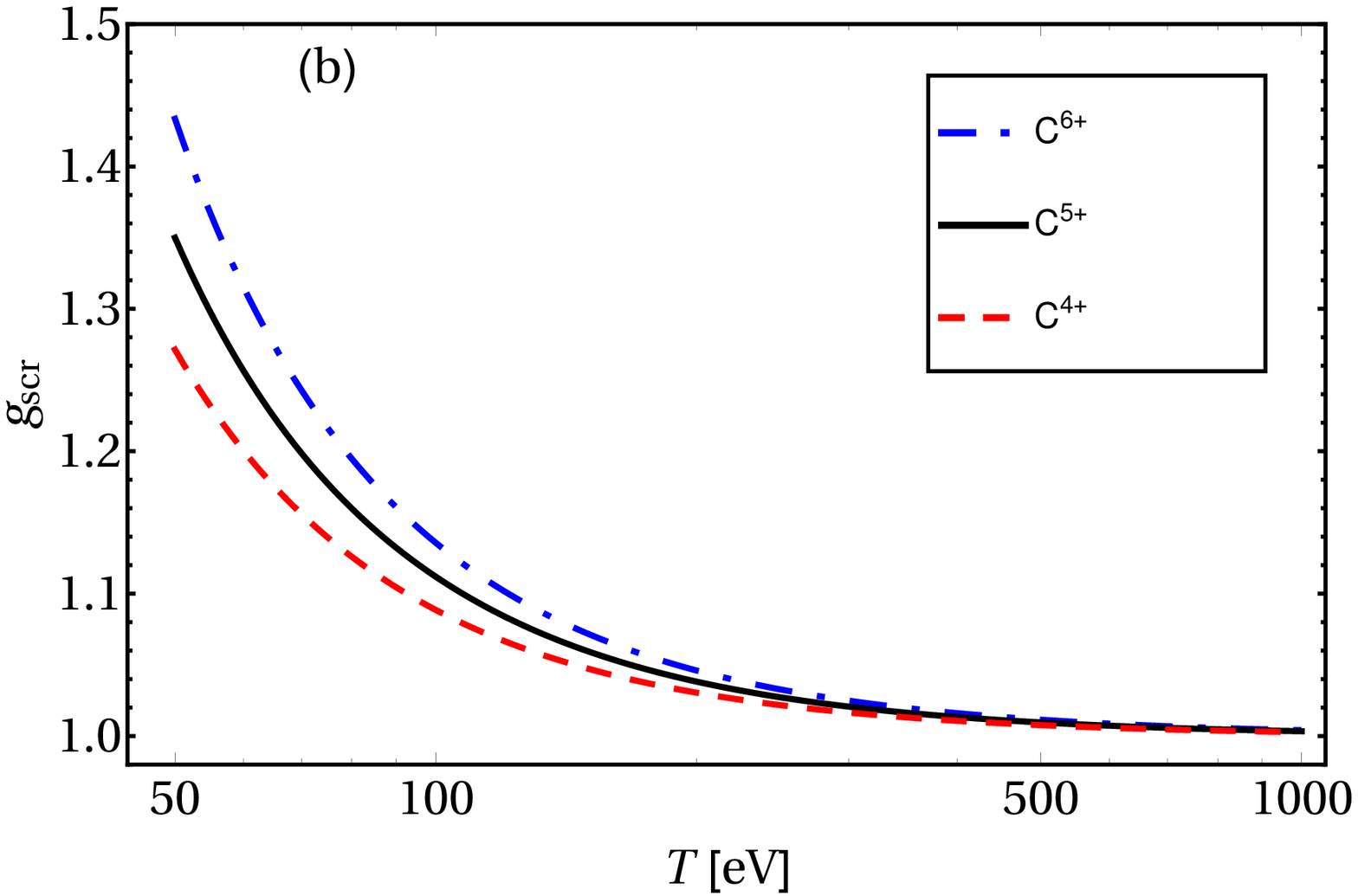}
\end{center}
\caption{Enhancement factor $g_{\rm{scr}}$ as a function of the temperature $T$ for the reaction $^{13}$C($^4$He, $n$)$^{16}$O in $^4$He plasmas for (a) the cases of C$^{6+}$ with different helium densities and (b) for the cases of $n_{\rm{He}} = 5\times 10^{20}$ cm$^{-3}$ with different carbon charge states.} \label{fig:screenfac}
\end{figure}

Figure \ref{fig:screenfac} shows the enhancement factor $g_{\rm{scr}}$ as a function of the temperature $T$ for the reaction $^{13}$C($^4$He, $n$)$^{16}$O in $^4$He plasmas predicted by the formula (\ref{eq:salscr}); results for selected helium densities are shown in Fig.~\ref{fig:screenfac}(a) and results for different carbon charge states are shown in Fig.~\ref{fig:screenfac}(b). The plots show that the enhancement factor approaches unity for high plasma temperatures. The screening effects are quite small for temperatures $T \sim 1$ keV in the plasmas under consideration. When decreasing the plasma temperature to $T \sim 100$ eV, the enhancement factor $g_{\rm{scr}}$ becomes significantly large because it is comparable to values in astrophysical scenarios [see, e.g., \cite{SalpeterAJP1954, DzitkoAPJ1995, GruzinovAPJ1998}]. Therefore, in the following, we focus  on the cases for $^4$He plasmas with temperatures of $T = 100$ eV.

\begin{figure}[h]
\begin{center}
\includegraphics[width=0.95\columnwidth]{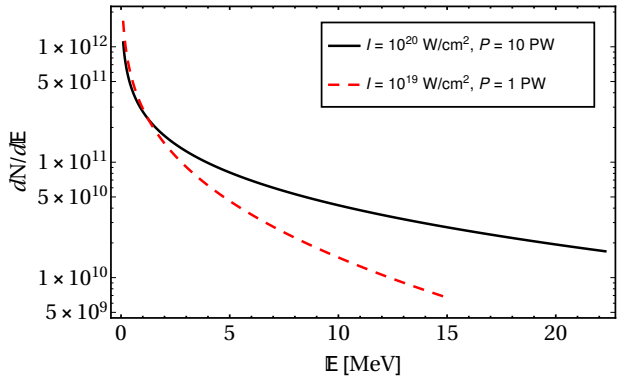}
\end{center}
\caption{Number of TNSA-accelerated $^{13}$C ions per unit energy. The thickness of the $^{13}$C target is $5$ $\rm{\mu m}$. Solid black line: laser parameters corresponding to case A described in text. In this case, the charge state of the $^{13}$C ions is $6+$. Dashed red curve:  laser parameters corresponding to case B described in the text. In this case, the charge state of the $^{13}$C ions is $4+$.} \label{fig:carbonspec}
\end{figure}

Turning to the $^{13}$C ions produced in the TNSA scenario, we show in Fig.~\ref{fig:carbonspec} two examples of ion spectra predicted by the TNSA model described in Sec.~\ref{sec:TNSA}. In these two examples, a $^{13}$C target  thickness of $5$ $\rm{\mu m}$ is considered. We consider here isotopically enriched $^{13}$C targets [enriched $^{13}$C targets are available and  have been already used in $^{13}$C-related experiments \citep{HarissopulosPRC2005, HeilPRC2008}]. Furthermore, since the ion beam from TNSA is originating from the rear surface of the target \citep{HegelichPRL2002, FuchsNP2006, MacchiRMP2013}, it is possible to obtain similar $^{13}$C ion beams using conductor targets covered with a thin $^{13}$C layer on the rear surface. We first consider the case of a laser of  peak intensity $I = 10^{20}$ $\rm{W/cm^{2}}$,  peak power $P = 10$ PW and  pulse duration $\tau_{\rm{laser}} = 25$ fs (FWHM) interacting with the target (we denote this case as case A). The space-charge field created at the rear surface of the target can reach a value higher than $10^{13}$ $\rm{V/m}$. Therefore, $^{13}$C ions with the charge state $6+$ can be produced. As a second case, we consider the interaction with a laser of  peak intensity $I = 10^{19}$ $\rm{W/cm^{2}}$,  peak power $P = 1$ PW, and pulse duration $\tau_{\rm{laser}} = 250$ fs (FWHM); we denote this case as case B. Now only a lower space-charge field can be created. Therefore, only up to C$^{4+}$ can be produced in this case. We adopt in this work a wavelength of $800$ nm for the laser, which is the same as the one of the two ELI-NP lasers \citep{RothELINP2015}. 

In Fig.~\ref{fig:carbonspec} it is shown that for case A a larger number of high-energy ions can be obtained than in case B, though they have similar values for the low-energy part. As shown in the ponderomotive scaling (\ref{eq:tepon}), the electron temperature is higher for the case of the high-intensity laser. Therefore, the exponential term in Eq.~(\ref{eq:enspe}) decreases more slowly for the high-intensity laser when increasing the ion energy. In other words, since the space-charge field created at the rear surface of the target is higher for case A, it is obvious that this configuration can accelerate more ions to high energy.

In order to obtain a pure carbon ion beam from TNSA, one needs to remove the contaminants that are always adsorbed on the target surface. This can be made by resistive heating [see \cite{HegelichPRL2002}] or surface cleaning [see \cite{AllenPRL2004}]. We note also that TNSA will produce multiple charge states simultaneously [see, e.g., \cite{HegelichPRL2002}]. As a first approximation, here we restrict our calculations to the highest charge state that can be reached only, as detailed in Sec.~\ref{sec:TNSA}.

Now we turn to the calculation of neutron events for the reaction $^{13}$C($^4$He, $n$)$^{16}$O in the envisaged experimental setup, considering the ion spectra calculated by the TNSA model described in Sec.~\ref{sec:TNSA}. In order to do this, knowledge of the nuclear cross-section for the reaction $^{13}$C($^4$He, $n$)$^{16}$O is also necessary. A combination of experimental results and a fit of the data for the reaction $^{13}$C($^4$He, $n$)$^{16}$O has been shown in \cite{XuNPA2013}. We adopt here, as an approximation, the fit of the $S(\mathbb{E}_{\rm{c}})$ in \cite{XuNPA2013} for the low-energy part ($\mathbb{E}_{\rm{c}} < 0.97$ $\rm{MeV}$), and the cross-section data form \cite{HarissopulosPRC2005} for the high-energy part ($\mathbb{E}_{\rm{c}} >0.97$ $\rm{MeV}$), respectively.

Taking into account that the density of ions from TNSA is normally much lower than the density of the secondary target under consideration, the influence of the fast ions from TNSA on the secondary target is expected to be negligible. The reaction events per laser pulse per unit energy of the incident fast ions are then given as
\begin{align}
   \frac{d \mathbb{N}}{d \mathbb{E}} &= \frac{d N}{d \mathbb{E}} g_{\rm{scr}} <\!\!\!\sigma(\mathbb{E}_{\rm{c}})\!\!\!> n_{\rm{s}} d,\label{eq:neutronplas} \\
   \frac{d \mathbb{N}}{d \mathbb{E}} &= \frac{d N}{d \mathbb{E}} \sigma(\mathbb{E}_{\rm{c}}) n_{\rm{s}} d, \label{eq:neutrongas}   
\end{align}
for fusion reactions in a secondary plasma target [Eq.~(\ref{eq:neutronplas})] and in a secondary gas target [Eq.~(\ref{eq:neutrongas})], respectively. Here $n_{\rm{s}}$ is the ion (atom) number density of the secondary target and $d$ is its thickness. Furthermore, $<\!\!\!\sigma(\mathbb{E}_{\rm{c}})\!\!\!>$ stands for the averaged cross-section under the Maxwellian distribution of the plasma target. For the sake of self-consistency and simplicity, in numerical calculations, $<\!\!\!\sigma(\mathbb{E}_{\rm{c}})\!\!\!>$ is obtained from a one-dimensional (1D) system, since the model adopted here for TNSA is also a 1D model.

It is worth mentioning here that the reaction events Eqs.~(\ref{eq:neutronplas}) and (\ref{eq:neutrongas}) are given under the assumption that the secondary target is a thin target such that the stopping of the fast ions passing through the target is negligible. Applying now the expressions above to the reaction $^{13}$C($^4$He, $n$)$^{16}$O in the considered experimental setup, $d\mathbb{N}/d \mathbb{E}$ of Eqs.~(\ref{eq:neutronplas}) and (\ref{eq:neutrongas}) give the neutron events per laser pulse per unit energy of the incident fast ions, and the density $n_{\rm{s}}$ becomes the $^4$He density $n_{\rm{He}}$.

\begin{figure}[h]
\begin{center}
\includegraphics[width=0.9\columnwidth]{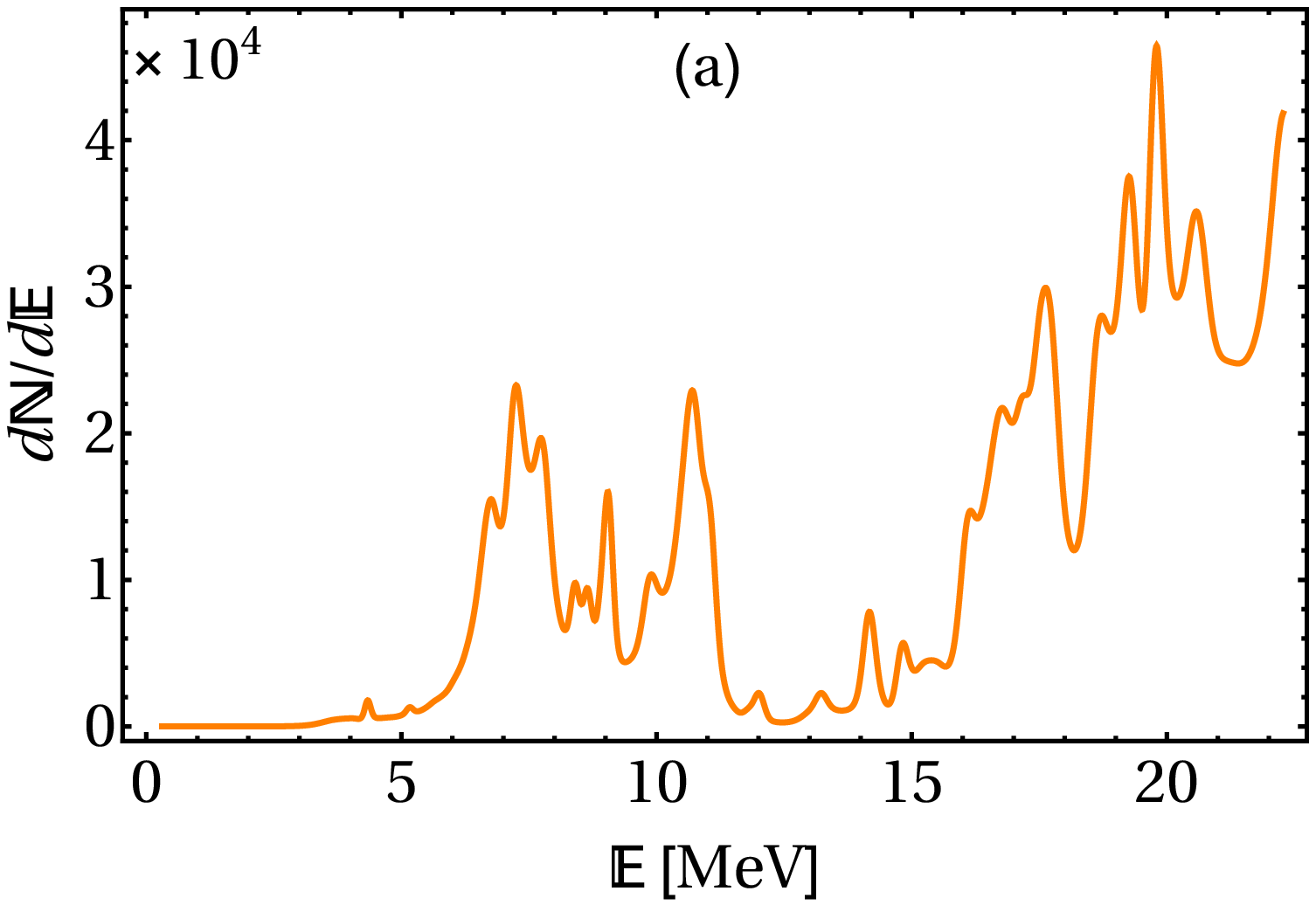}
\includegraphics[width=0.9\columnwidth]{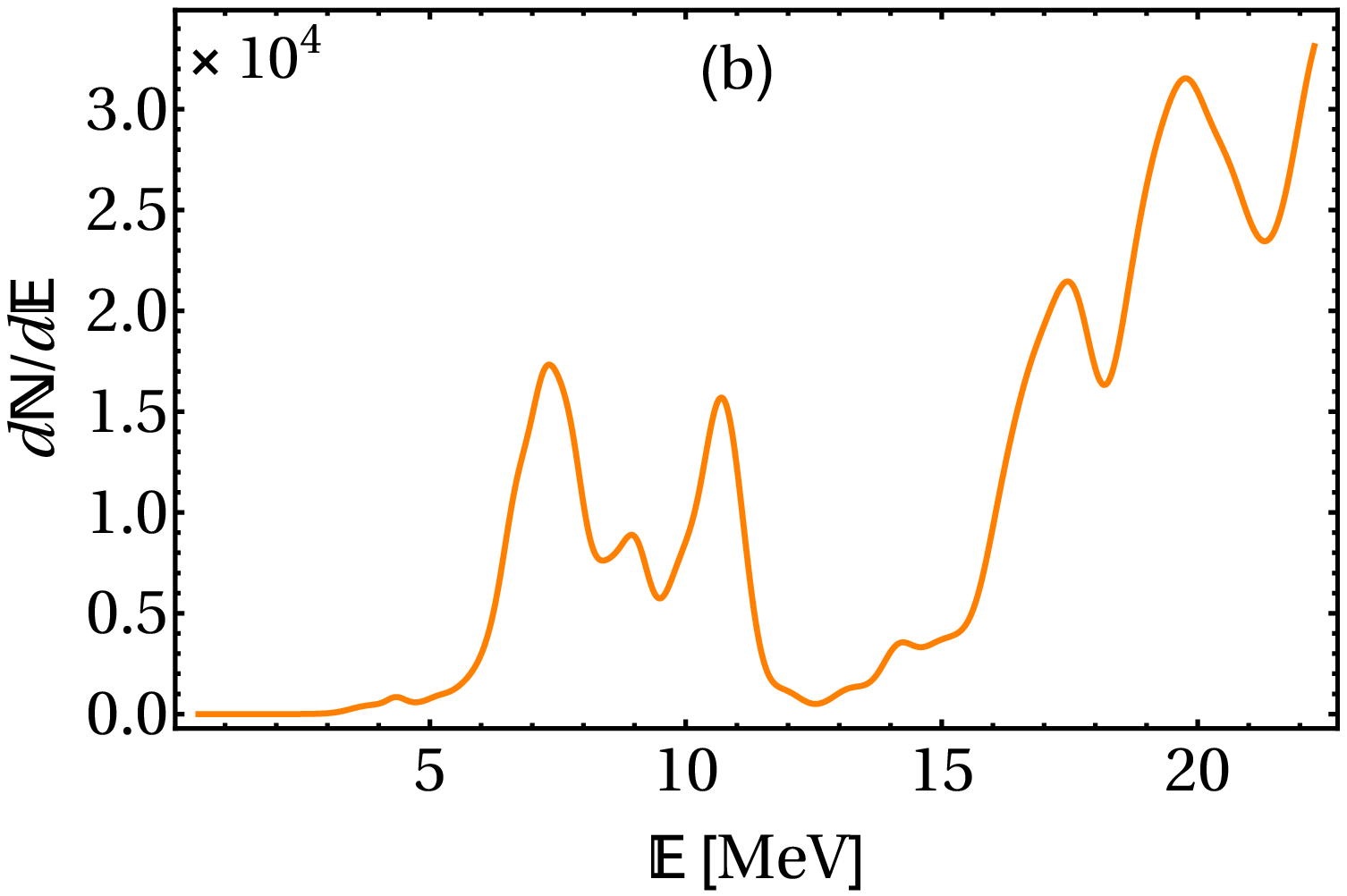}
\end{center}
\caption{Number of neutron events per laser pulse per unit energy of the incident $^{13}$C ions interacting with a plasma target. For the TNSA part, the thickness of the $^{13}$C target is $5$ $\rm{\mu m}$, the peak intensity of the incident  laser is $10^{20}$ $\rm{W/cm^{2}}$, the peak power of the laser pulse is $10$ PW, and the pulse duration is $25$ fs, respectively (case A discussed in the text). We consider the $^4$He plasma density $n_{\rm{He}}$ $5 \times 10^{20}$ cm$^{-3}$,  the thickness  $100$ $\rm{\mu m}$ and the temperatures (a) $100$ eV and (b) $1$ keV, respectively.} \label{fig:neutronspec}
\end{figure}

We show in Fig.~\ref{fig:neutronspec} the neutron spectra for the reaction $^{13}$C($^4$He, $n$)$^{16}$O in $^4$He plasmas with temperatures  of $T = 100$ eV and $T = 1$ keV. For the TNSA part, we adopt the same target and laser parameters as in the case A of Fig.~\ref{fig:carbonspec}. The $^4$He plasma density $n_{\rm{He}}$ is $5 \times 10^{20}$ cm$^{-3}$ and the thickness of the plasma is $100$ $\rm{\mu m}$. We can learn from Fig.~\ref{fig:neutronspec} that when the $^{13}$C ion energy $\mathbb{E} \lesssim 4$ MeV, the number of neutron events is very small. This is mainly due to the small cross-section of the reaction for $\mathbb{E}_{\rm{c}} \lesssim 1$ MeV. Figure \ref{fig:neutronspec} also shows a quite strong averaging effect due to the plasma for the case of $T = 1$ keV. The feature of detailed resonances in the cross-section [see \cite{XuNPA2013}] is washed out by the plasma at the higher temperature, and the neutron spectrum has thus a smoother appearance. In contrast, for the case of $T = 100$ eV shown in Fig.~\ref{fig:neutronspec}(a), the neutron spectrum clearly follows the detailed feature of the cross-section.

Integrating Eqs.~(\ref{eq:neutronplas}) and (\ref{eq:neutrongas}), we obtain the total number $\mathbb{N}$ of neutron events per laser pulse for the reaction under investigation. The results of the total number $\mathbb{N}$ of neutron events per laser pulse are shown in Table \ref{tabneutrons}, in the cases of $^4$He plasma with density $n_{\rm{He}} = 5 \times 10^{20}$ cm$^{-3}$, thickness $100$ $\rm{\mu m}$, and temperature $100$ eV. Several laser parameter sets and target thicknesses for TNSA are adopted in Table \ref{tabneutrons}. For the sake of comparison, we also show in Table \ref{tabneutrons} the total number of neutron events for the cases of a $^4$He gas target. The number of neutron events in the case of the carbon ions interacting with the plasma is significantly larger than the one interacting with the gas target; in particular, the differences are larger than $10\%$ for the cases of C$^{6+}$. These differences are mainly coming from the screening enhancement, since the energy contribution from the plasma to the reaction is negligible for the case of $T = 100$ eV [see also Fig.~\ref{fig:neutronspec}(a)]. The detection of these differences  could determine for the first time experimentally  the screening enhancement factor.

\begin{table}[h]  \addtolength{\tabcolsep}{-1.25pt}
\begin{center}
\begin{tabular}{lcccccc}
  \hline\hline
  $P$ [PW]                                        & $10$         & $10$         &  $10$        & $10$         & $5$                       & $1$ \\
  $I$ [$\rm{W/cm^{2}}$]                    & $10^{20}$ & $10^{20}$ & $10^{20}$ & $10^{20}$ & $3\times 10^{19}$ & $10^{19}$ \\
  $\tau_{\rm{laser}}$ [fs]                   & $25$          & $25$         & $25$         & $25$         & $50$                     & $250$  \\
  $d_{\rm{t}}$ [$\rm{\mu m}$]           & $10$          &  $5$          &  $1$          & $0.1$        & $5$                       & $5$ \\
  \hline
  $^{13}$C ion charge state                   & $6+$    & $6+$    & $6+$   & $6+$     & $5+$   & $4+$ \\
  $\mathbb{N}$ for plasma [$10^4$]      & $19.6$ & $24.5$ & $30.1$ & $31.5$ & $5.69$ & $2.75$ \\
  $\mathbb{N}$ for gas target [$10^4$]  & $17.3$ & $21.6$ & $26.5$ & $27.7$ & $5.12$ & $2.53$ \\
   \hline\hline
\end{tabular}
\end{center}
\caption{The total number $\mathbb{N}$ of neutron events per laser pulse for $^4$He gas and plasma targets. We consider the target density $n_{\rm{He}} = 5 \times 10^{20}$ cm$^{-3}$ and target thickness $100$ $\rm{\mu m}$ for both cases. For the $^4$He plasma target, the plasma temperature is $100$ eV. Results for several laser parameter sets and target thicknesses for TNSA are presented.}
\label{tabneutrons}
\end{table}

Furthermore, we show in Table \ref{tabndensities} the total number of neutron events per laser pulse for selected $^4$He densities. Here, we adopt the same target and laser parameters as in case A of Fig.~\ref{fig:carbonspec} for the TNSA part. As expected, the difference between the two scenarios of fast ions interacting with the plasma and gas targets increases with increasing density of the $^4$He target. As shown in Table \ref{tabndensities}, the difference reaches $\sim 20\%$ for the density $n_{\rm{He}} = 10^{21}$ cm$^{-3}$. Such high-density gas jet targets are experimentally more challenging to realize, but have been recently demonstrated \citep{SyllaRSI2012}.

\begin{table}[h]  \addtolength{\tabcolsep}{5pt}
\begin{center}
\begin{tabular}{lcccccc}
  \hline\hline
  $n_{\rm{He}} $ [cm$^{-3}$]  & $10^{21}$ & $5\times 10^{20}$ &  $10^{20}$  \\
  \hline
  $\mathbb{N}$ for plasma [$10^4$]     & $51.6$ & $24.5$ & $4.57$ \\
  $\mathbb{N}$ for gas target [$10^4$] & $43.1$ & $21.6$ & $4.31$ \\
   \hline\hline
\end{tabular}
\end{center}
\caption{The total number $\mathbb{N}$ of neutron events per laser pulse for $^4$He gas and plasma targets. The thickness of the $^4$He target is in both cases $100$ $\rm{\mu m}$. For the plasma target, the temperature is $100$ eV. In the TNSA part, the thickness of the $^{13}$C target is $5$ $\rm{\mu m}$, the peak intensity of the incident  laser is $10^{20}$ $\rm{W/cm^{2}}$,  the peak power of the laser pulse is $10$ PW, and the pulse duration is $25$ fs, respectively (case A discussed in the text).}
\label{tabndensities}
\end{table}

\begin{table}[h]  \addtolength{\tabcolsep}{-2.2pt}
\begin{center}
\begin{tabular}{lcccccc}
  \hline\hline
  $P$ [PW]                                        & $10$         & $5$                        &  $1$          & $10$        & $5$                        & $1$ \\
  $I$ [$\rm{W/cm^{2}}$]                    & $10^{20}$ & $3\times 10^{19}$ & $10^{19}$ & $10^{20}$ & $3\times 10^{19}$ & $10^{19}$ \\
  $\tau_{\rm{laser}}$ [fs]                   & $25$          & $50$                     & $250$        & $25$         & $50$                     & $250$  \\
  $d_{\rm{t}}$ [$\rm{\mu m}$]           & $5$            &  $5$                      &  $5$           & $5$           & $5$                       & $5$ \\
  \hline
  $^{13}$C ion charge state           & $6+$    & $5+$    & $4+$   & $6+$     & $5+$   & $4+$ \\
  \hline
  $T$ [keV]                  & $0.5$   & $0.5$   & $0.5$   & $1$      & $1$     & $1$ \\
  $\mathbb{N}$ for plasma [$10^4$]     & $21.8$ & $5.13$ & $2.55$ & $21.7$ & $5.06$ & $2.54$ \\
  \hline
  $\mathbb{N}$ for gas target [$10^4$] & $21.6$ & $5.12$ & $2.53$ & $21.6$ & $5.12$ & $2.53$ \\
   \hline\hline
\end{tabular}
\end{center}
\caption{The total number $\mathbb{N}$ of neutron events per laser pulse for selected temperatures $T$ of the $^4$He plasma target, and for the case of a gas target. The density of the $^4$He target all cases is $n_{\rm{He}} = 5 \times 10^{20}$ cm$^{-3}$ and its thickness is $100$ $\rm{\mu m}$, respectively.}
\label{tabntemp}
\end{table}

We also study the effects of the plasma temperature on the total number of neutron events. The results of the number $\mathbb{N}$ of neutron events per laser pulse for selected temperatures $T$ of $^4$He plasmas are shown in Table \ref{tabntemp}. The values show that for the plasma temperatures $500$ eV and $1$ keV, the differences between the two scenarios of fast ions interacting with the plasma target and the gas target are much smaller than the ones shown in Tables \ref{tabneutrons} and \ref{tabndensities}. The screening effects are very small for high plasma temperatures at the densities considered. Furthermore, with increasing plasma temperature, the energy contribution from the plasma to the reaction center-of-mass energy grows. This will also effect the number of neutron events. Therefore, as shown in Table \ref{tabntemp}, for the case of C$^{5+}$ and $T = 1$ keV, the number of neutron events in the plasma can be even smaller than the one in the gas target.

\section{Summary and conclusions \label{sec:sumcon}} 

We have theoretically investigated a proposed experimental setup for nuclear reactions in laser-generated plasmas at ELI-NP, where two laser beams generate two colliding plasmas \citep{RothELINP2015}. One of these two plasmas is a rapidly streaming plasma (ion flow), which is produced by an intense laser pulse interacting on a solid target through the TNSA mechanism. The second plasma in this experimental setup is a cold plasma that is produced by the interaction of a second laser pulse on a gas jet target. We have used an isothermal, fluid model \citep{MoraPRL2003, FuchsNP2006} to model the TNSA scheme. With the ion spectrum obtained from this model, we have studied the fusion reactions in the interaction of the rapidly streaming plasma (ion flow) on the secondary target.

As the first case of study, we have considered the reaction $^{13}$C($^4$He, $n$)$^{16}$O, which is one of the important helium burning processes, as well as one of the main neutron sources for the $s$-process \citep{LooreBook1992, GallinoAPJ1998, HeilPRC2008}. We have obtained the neutron spectra and the total number of neutron events expected in this experimental setup, based on the ELI-NP laser parameters of \cite{RothELINP2015}. We have found that it is possible to determine the plasma screening enhancement factor \citep{SalpeterAJP1954, GruzinovAPJ1998} for fusion reactions in plasmas by detecting the difference of the reaction events between the two scenarios of ion flow interacting with the plasma target and the gas target (i.e., without the use of the second laser pulse). By selecting appropriate experimental conditions, this difference in neutron events for reaction $^{13}$C($^4$He, $n$)$^{16}$O exceeds $10\%$ and can even reach $\sim 20\%$. This work shows a possible way to experimentally access and understand the plasma screening effect for nuclear reactions in stellar environments in the weak-screening regime. This could allow for the first time the evaluation of nuclear reactions in stellar environments. While it is obvious that laser-generated plasmas in the laboratory cannot cover all parameter ranges present in astrophysical environments, a verification of the screening theory in the available weak-screening regime would  be very valuable and significantly advance the field of nuclear astrophysics.

\acknowledgments{We thank C. H. Keitel and S. Tudisco for fruitful discussions.}

\bibliography{snrpbib}

\end{document}